\documentclass[12pt]{iopart}
\usepackage[pdftex]{graphicx}
\usepackage{iopams}
\usepackage{cite}

\newcommand{\bc}{\begin{center}}
\newcommand{\ec}{\end{center}}
\begin{document}

\title{How can we check the uncertainty relation?}

\author{Vladimir N. Chernega}

\address{P. N. Lebedev Physical Institute, Leninskii
Prospect 53, Moscow 119991, Russia}

\ead{vchernega@gmail.com}

\begin{abstract}
The state-extended uncertainty relations generalizing the Robertson
Schr\"odinger inequality are presented in the form appropriate for
the experimental check by homodyne photon state detection. The
method of qubit portrait of qudit states identified with the
tomographic probability distributions is discussed to analyze the
entanglement of two-mode field.
\end{abstract}

\pacs{42.50.-p, 42.50.Dv, 03.67.-a}


\section{Introduction}
There are recent results where the authors discuss the quantum
theory going beyond conventional quantum mechanics
\cite{Sweinberoon,t'Moof,Peres}. In this connection the precise
experimental check of basic quantum phenomena with high accuracy,
e.g. quantum uncertainty relations like Heisenberg
position--momentum uncertainty relation \cite{Heis27}, Robertson
\cite{Rob27} and Shr\"odinger \cite{Schr27} uncertainty relations,
purity dependent uncertainty relations \cite{Dodman183,Dod} and
other quantum inequalities would be interesting to fulfill. The new
formulation of quantum mechanics based on tomographic probability
representation of quantum states
\cite{Ibort,ManciniPLA1996,M.A.FoundPhys} provides convenient tools
to suggest such experiments \cite{Adv.SciLettMarmo} using the
homodyne photon state detection where optical tomograms of the
photon quantum states are measured \cite{SolimenoPorxio}. In
\cite{Trifonov1,2} new quantum uncertainty relations were found. In
contrast to Heisenberg and Robertson-Schr\"odinger uncertainty
relations Trifonov inequalities depend on two quantum states and
they were called state-extended uncertainty relations. In fact these
inequalities provide a generalization of standard position-momentum
uncertainty relations. The state-extended generalization of
Heisenberg uncertainty relations was studied in \cite{ChernegaJRLR}
and its tomographic form was found and proposed for experimental
check in the photon homodyne detection. In this work we consider
another state-extended generalization of position-momentum
uncertainty relations. Our aim is to obtain tomographic form of
Trifonov inequality which is state-extended Robertson-Schr\"odinger
uncertainty relations containing covariance of position and
momentum. Another our goal is to review probability representation
approach in the context of studying qubit portrait of qudit states
\cite{ChernegaManJRLR,CosmoLupoManJPA}. The paper is organized as
follows. In next section 2 we present the optical tomography scheme
of one-mode quantum electromagnetic field. (see, e.g. \cite{Ibort}).
In Sec.3 we give Trifonov inequalities in tomographic form. In Sec.4
we discuss qubit portrait method and in Sec.5 present conclusions
and prospects.

\section{Optical tomography}
The quantum state in probability representation of quantum mechanics
is determined by optical tomogram $w(X,\theta)$. Here
$-\infty<X<+\infty$, $0\leq\theta\leq 2\pi$. The optical tomogram is
probability density of random homodyne quadrature $X$. It depends on
the angle $\theta$ which in quantum optics is called local
oscillator phase. The optical tomogram provides the density operator
$\hat{\rho}$ of the photon quantum state
\begin{eqnarray}
\hat{\rho}=\frac{1}{2\pi}\int\limits_0^{\pi}d\theta\int\limits^{+\infty}_{-\infty}d\eta
dX\omega(X,\theta)|\eta|
\exp{i\eta(X-\hat{q}\cos\theta-\hat{p}\sin\theta)}.
\end{eqnarray}\label{eq.1.1}
The optical tomogram can be found if the density operator
$\hat{\rho}$ is known
\begin{eqnarray}
w(X,\theta)=\mbox{Tr}\,\hat\rho\delta(X-\cos\theta\hat
q-\sin\theta\hat p).\label{eq.1.2}
\end{eqnarray}
The physical meaning of the optical tomogram is the following one.
It is nonnegative probability density of the homodyne quadrature
\begin{eqnarray}
X=\hat{q}\cos\theta+\hat{p}\sin\theta.
\end{eqnarray}\label{eq.1.2}
Consequently for $\theta=0$ the tomogram in quantum optics provides
the probability distribution of first quadrature $q$ and for
$\theta=\pi/2$ the tomogram yields the probability distribution of
the second quadrature $p$. In quantum mechanics for $\theta=0$ and
$\theta=\pi/2$ the tomogram provides probability distributions of
position and momentum, respectively. The most important property of
the optical tomogram is that it is measured experimentally
\cite{RaymerPRL,RaymerLvovsky,Bellini}. For pure state with wave
function $\Psi(y)$ the tomogram reads
\begin{eqnarray}\label{eq.1.4}
w(X,\theta)=\frac{1}{2\pi
|\sin\theta|}\left|\int\Psi(y)\exp\left(\frac{iy^2}{2\tan\theta}-\frac{i
X y}{\sin\theta}\right)\,dy\right|^2.
\end{eqnarray}
If the Hamiltonian $\hat{H}=\hat{p}^2/2+U(\hat{q})$, the optical
tomogram obeys the evolution equation of the form
\cite{KorenJRLR2011}
\begin{eqnarray}
\fl \left.\frac{\partial}{\partial t}\right. w(X,\theta,t)
=\left[\cos^2\theta\frac{\partial}{\partial\theta}-\frac{1}{2}\sin
2\theta
\left(1+X\frac{\partial}{\partial X}\right)\right]w(X,\theta,t)\nonumber\\
\fl
+\frac{1}{i}\left\{V\left[\left(\sin\theta\frac{\partial}{\partial\theta}\left(\frac{\partial}{\partial
X}
\right)^{-1}+X\cos\theta +i\frac{\sin\theta}{2}\frac{\partial}{\partial X}\right)\right]\right.\nonumber\\
\fl
-\left.V\left[\left(\sin\theta\frac{\partial}{\partial\theta}\left(\frac{\partial}{\partial
X} \right)^{-1}+X\cos\theta
-i\frac{\sin\theta}{2}\frac{\partial}{\partial
X}\right)\right]\right\} w(X,\theta,t).\label{eq.1.11}
\end{eqnarray}

\section{Uncertainty relations}
In view of physical meaning of the optical tomogram the Heisenberg
uncertainty relation
\begin{eqnarray}\label{eq.1.5}
\sigma_{qq}\sigma_{pp}\geq {1}/{4}
\end{eqnarray}
can be presented in the tomographic form as \cite{Olga}
\begin{eqnarray}\label{eq.1.10}
\left[\int w(X,0)X^2\,d X-\left(\int
w(X,0)X\,dX\right)^2\right]\nonumber\\ \times\left[\int
w(X,\pi/2)X^2\,dX\right.-\left.\left(\int
w(X,\pi/2)X\,dX\right)^2\right]\geq \frac{1}{4}\,.
\end{eqnarray}
The Robertson-Schr\"odinger inequality
\begin{eqnarray}\label{eq.1.12}
\sigma_{qq}\sigma_{pp}-\sigma^2_{qp}\geq {1}/{4}.
\end{eqnarray}
where quadrature variances and covariance are calculated for the
same state was generalized by Trifonov \cite{Tr2001,2}. For two pure
states $|\Psi_1\rangle$,$|\Psi_2\rangle$ this state-extended
inequality reads
\begin{eqnarray}
\frac{1}{2}\left[\mbox{Tr}\left(\hat{q}^2|\Psi_1\rangle\langle\Psi_1|\right)
-(\mbox{Tr}\left(\hat{q}|\Psi_1\rangle\langle\Psi_1|\right))^2\right]\nonumber\\
\times\left[\mbox{Tr}\left(\hat{p}^2|\Psi_2\rangle\langle\Psi_2|\right)-
(\mbox{Tr}\left(\hat{p}|\Psi_2\rangle\langle\Psi_2|\right))^2\right]\nonumber\\
+\frac{1}{2}\left[\mbox{Tr}\left(\hat{q}^2|\Psi_2\rangle\langle\Psi_2|\right)
-(\mbox{Tr}\left(\hat{q}|\Psi_2\rangle\langle\Psi_2|\right))^2\right]\nonumber\\
\times
\left[\mbox{Tr}\left(\hat{p}^2|\Psi_1\rangle\langle\Psi_1|\right)-
(\mbox{Tr}\left(\hat{p}|\Psi_1\rangle\langle\Psi_1|\right))^2\right]\nonumber\\
-\left\{\mbox{Tr}\left(\frac{\hat{q}\hat{p}+\hat{p}\hat{q}}{2}|\Psi_2\rangle\langle\Psi_2|\right)-
\mbox{Tr}\left(\hat{q}|\Psi_2\rangle\langle\Psi_2|\right)
\mbox{Tr}\left(\hat{p}|\Psi_2\rangle\langle\Psi_2| \right)\right\}\nonumber\\
\times\left\{\mbox{Tr}\left(\frac{\hat{q}\hat{p}+\hat{p}\hat{q}}{2}|\Psi_1\rangle\langle\Psi_1|\right)-
\mbox{Tr}\left(\hat{q}|\Psi_1\rangle\langle\Psi_1|\right)
\mbox{Tr}\left(\hat{p}|\Psi_1\rangle\langle\Psi_1| \right)\right\}
\geq \frac{1}{4}.\label{eq.1.12}
\end{eqnarray}
This inequality can be written in the tomographic form and it reads
\begin{eqnarray}
&&\frac{1}{2}\left[\int w_1(X,\theta)X^2\,d X-\left(\int
w_1(X,\theta)X\,d X\right)^2\right]\nonumber\\
&&\times\left[\int w_2(X,\theta+\pi/2)X^2\,d X-\left(\int
w_2(X,\theta+\pi/2)X\,dX\right)^2\right]
\nonumber\\
&&+\frac{1}{2}\left[\int w_2(X,\theta)X^2\,d X-\left(\int
w_2(X,\theta)X\,dX\right)^2\right]\nonumber\\
&&\times\left[\int w_1(X,\theta+\pi/2)X^2\,d X-\left(\int
w_1(X,\theta+\pi/2)X\,dX\right)^2\right]\nonumber\\
&&-\{\int w_1(X,\theta+{\pi\over 4})X^2\,d X-\left(\int
w_1(X,\theta+{\pi\over 4})X\,d
X\right)^2\,\nonumber\\
&&-\frac{1}{2}\left[\int w_1(X,\theta)X^2\,d X-\left(\int
w_1(X,\theta)X\,d X\right)^2\right]\nonumber\\
&&\,-\frac{1}{2}\left[\int w_1(X,\theta+{\pi\over2})X^2\,d
X-\left(\int
w_1(X,\theta+{\pi\over2})X\,d X\right)^2\right]\}\nonumber\\
&&\times\{\int w_2(X,\theta+{\pi\over 4})X^2\,d X-\left(\int
w_2(X,\theta+{\pi\over 4})X\,d
X\right)^2\,\nonumber\\
&&-\frac{1}{2}\left[\int w_2(X,\theta)X^2\,d X-\left(\int
w_2(X,\theta)X\,d X\right)^2\right]\nonumber\\
&&\,-\frac{1}{2}\left[\int w_2(X,\theta+{\pi\over2})X^2\,d
X-\left(\int w_2(X,\theta+{\pi\over2})X\,d X\right)^2\right]\}
\geq{1}/{4}.\label{eq.1.13}
\end{eqnarray}
The obtained inequalities can be checked if both tomograms
$w_1(X,\theta)$ and $w_2(X,\theta)$ are measured. This inequality
takes place for mixed state too.

\section{Qubit portrait and inequalities for optical tomograms}
Qubit portrait of qudit states provides the probability distribution
given by two positive numbers $p_1,p_2$, where $p_1+p_2=1$ obtained
from an initial probability distribution ${\cal P}_1, {\cal
P}_2\ldots,{\cal P}_N$ where $\sum_k{\cal P}_k=1$. This qubit
probability distribution can be obtained using linear map of
$N$-vector with components ${\cal P}_k$ onto two-vector with
components $p_1,p_2$. Th map can be described e.g. by the
corresponding stohastic matrix. If one has probability density
$w(X,\theta)$ the qubit portrait can be also constructed  by using
the rectangular matrix
\begin{equation}\label{2.1}
p_m(\theta)=\int K_m(X) w(X,\theta)d X, \quad
m=\frac{1}{2},-\frac{1}{2}
\end{equation}
where $w(X,\theta)$ is the tomogram of a quantum state. If one has
two-mode state with the optical tomogram
$w(X_1,X_2,\theta_1,\theta_2)$ the generalized qubit portrait
provides the analog of spin-tomogram of two-qubits
\begin{equation}\label{2.2}
p(m_1,m_2,\theta_1,\theta_2)=\int
K_{m_1m_2}(X_1,X_2)w(X_1,X_2,\theta_1,\theta_2)d X_1 d X_2.
\end{equation}
For example the matrix $K_{m_1m_2}(X_1, X_2)$ can have factorized
form. The quantum correlations for the two-mode states can be
studied by considering the properties of the function (\ref{2.2}).
For example the probability four-vector $\vec{p}(\theta_1,\theta_2)$
depending on extra angle parameters can be studied analogously to
the case of studying spin tomographic probability of two-qubit
entangled state for which the Bell inequality violation is
sufficient condition of the state entanglement. Then the Bell number
is given in terms of the function $p(m_1,m_2,\theta_1,\theta_2)$ as
follow
\begin{eqnarray}
&&B=\mbox{max}|p_{+{1\over2}\,+{1\over2}}(\theta_1,\theta_2)
-p_{+{1\over2}\,-{1\over2}}(\theta_1,\theta_2)-
p_{-{1\over2}\,+{1\over2}}(\theta_1,\theta_2)+
p_{-{1\over2}\,-{1\over2}}(\theta_1,\theta_2)\nonumber\\
&&+p_{+{1\over2}\,+{1\over2}}(\theta_1,\theta_3)
-p_{+{1\over2}\,-{1\over2}}(\theta_1,\theta_3)-
p_{-{1\over2}\,+{1\over2}}(\theta_1,\theta_3)+
p_{-{1\over2}\,-{1\over2}}(\theta_1,\theta_3)\nonumber\\
&&+p_{+{1\over2}\,+{1\over2}}(\theta_4,\theta_2)
-p_{+{1\over2}\,-{1\over2}}(\theta_4,\theta_2)-
p_{-{1\over2}\,+{1\over2}}(\theta_4,\theta_2)+
p_{-{1\over2}\,-{1\over2}}(\theta_4,\theta_2)\nonumber\\
&&-p_{+{1\over2}\,+{1\over2}}(\theta_4,\theta_3)
+p_{+{1\over2}\,-{1\over2}}(\theta_4,\theta_3)+
p_{-{1\over2}\,+{1\over2}}(\theta_4,\theta_3)-
p_{-{1\over2}\,-{1\over2}}(\theta_4,\theta_3)|.\label{2.3}
\end{eqnarray}
For factorized matrix $K_{m_1}^{(1)}(X_1)K_{m_2}^{(2)}(X_2)$ the
violation of inequality $B\leq2$ is sufficient condition to conclude
that the two-mode state with the tomogram
$w(X_1,X_2,\theta_1,\theta_2)$ is entangled. It means that the
optical tomogram of such entangled state can not be presented in the
form of convex sum
\begin{equation}\label{2.4}
w(X_1,X_2,\theta_1,\theta_2)=\sum_k p_k
w_1^{(k)}(X_1,\theta_1)w_2^{(k)}(X_2,\theta_2)
\end{equation}
where $w_1^{(k)}(X_1,\theta_1)$ and $w_2^{(k)}(X_2,\theta_2)$ are
optical tomograms of the first and second mode states. These
tomograms must satisfy also the Trifonov inequality (\ref{eq.1.13}).

\section{Conclusions}
To conclude we point out the main results of our work. We derived
new inequalities for optical tomograms of quantum states which are
obtained from Trifonov state-extended inequalities and presented in
the form aprropriate for experimental check by means of homodyne
photon detection. We applied the recent qubit portrait method of
studying qudit states to introduce a method to analyze entanglement
of two-mode electromagnetic field state. Using the map of the
optical tomogram of two-mode state onto analog of the spin-tomogram
of two qubits we founs that the violaion of Bell inequalities for
the obtained analog of spin tomogram is sufficient condition for the
presence of the entanglement in the two-mode state under study.

\section*{Acknowledgements}

This study was partially supported by the Russian Foundation for
Basic Research under Projects Nos.~10-02-00312 and 11-02-00456. The
author is grateful to the Organizers of CEWQO-2011 (Universidad
Complutense, Madrid, Spain) for kind hospitality.

\end{document}